\documentclass{article}

\usepackage{arxiv}

\usepackage[utf8]{inputenc} % allow utf-8 input
\usepackage[T1]{fontenc}    % use 8-bit T1 fonts
\usepackage{hyperref}       % hyperlinks
\usepackage{url}            % simple URL typesetting
\usepackage{booktabs}       % professional-quality tables
\usepackage{amsfonts}       % blackboard math symbols
\usepackage{nicefrac}       % compact symbols for 1/2, etc.
\usepackage{microtype}      % microtypography
\usepackage{lipsum}
\usepackage{graphicx}
\usepackage{amsmath}
\usepackage{subfigure}
\usepackage{authblk}

\title{Exploring a New Model for Mobile Positioning Based on CDR Data of The Cellular Networks \\ {\small \textit{From Coverage Area to Positioning}} }

\author{
  Amnir Hadachi\thanks{The authors gratefully acknowledge the contribution of The Software Technology and Applications Competence Centre (STACC) through Large-scale Mobile Positioning Data Mining (Demograft) project and all the partners in Archimedes project "The Real-time Location-based Big Data Algorithms" for their help in providing the data. This research was supported by IUT34-4 "Data Science Methods and Applications" (DSMA) project.}  and Artjom Lind\\
    ITS Lab, Institute of Computer Science\\
    University of Tartu\\
    51014 Tartu, Estonia \\
  \texttt{hadachi@ut.ee} \\
}

\begin{document}
\maketitle

\begin{abstract}
The emerging technologies related to mobile data especially CDR data has great potential for mobility and transportation applications. However, it presents some challenges due to its spatio-temporal characteristics and sparseness. Therefore, in this article, we introduced a new model to refine the positioning accuracy of mobile devices using only CDR data and coverage areas locations. The adopted method has three steps: first, we discovered which model of movement (\textit{Move, Stay}) is associated with the coverage areas where the mobile device was connected using a Kalman filter. Then, simultaneously we estimated the location or the position of the device. Finally, we applied map-matching to bring the positioning to the right road segment. The results are very encouraging; nevertheless, there is some enhancement that can be done at the level of movement models and map matching. For example by introducing more sophisticated movement model based on data-driven modeling and a map matching that uses the movement model type detected by matching \textit{Stay} location to buildings and \textit{Move} model to roads. 
\end{abstract}

% keywords can be removed

\section{Introduction}
\label{intro}

The rapid development of information and communication technologies has made them very valuable for humanity. Since they are used as a supporting tool for organizing our daily life tasks and duties. It is clear that smartphones have a great potential of being the ultimate sensor that can reveal many aspects of human mobility and behavior \cite{Etienne2017}. From this perspective, much research has been done and many applications created in order to evaluate or demonstrate the great potential behind mobile data in depicting human mobility \cite{Shan2017}, road traffic status \cite{Noe2012}, tourists movement, and their displacement \cite{Ratul2016}. 

Understanding human mobility behavior is one of the major challenges in many research domains and applications. The reason behind this is its implication directly or indirectly in many aspects of our daily life, such as urban planning \cite{Hadachi2018}, traffic management\cite{Jane2012}, crowd analysis, and population displacement \cite{Lind2017}, targeted marketing or advertisement \cite{Yan2013},  monitoring, and control \cite{Fia2012}. After all, mobile phones are literally deployed sensors used as urban sensing devices, reporting on the status of our interactions with our cities, environment and between us \cite{Pinel2015}. 

Furthermore, mobile phones provide a rich and massive amount of data with its interesting information and challenges\cite{Lu2016}. The Call Detail Records (CDR) are the data collected by the mobile operator for billing purposes and it has been an interesting type of data investigated by researchers and also industrials to extract insights and knowledge about human mobility patterns \cite{Pinel2015}.  

CDR data contains information about the events recorded, their time of occurrence and the ID code of the antenna that provided the connection \cite{Ela2015}. These events are registered the moment the mobile device is in use for calling or messaging or browsing the internet. Hence, CDR data have this characteristic of being temporal data, but, it can be also considered spatio-temporal data since we can associate the coverage area shapes to the ID code of the antennas using the radio waves data.  

The fact that the coverage areas are a representation of radio waves rise up many issues related to the fluctuation of the signal strength due to many factors, such as weather, building, lakes, etc \cite{Doh2014}. Additionally, the CDR data is very sparse in time and space; because the events are registered only when the phone is in use and the coverage areas locations are very granular. Thus, CDR data has some potential since it can reflect human mobility but in order to do so there are some challenges and obstacle that has to be dealt in order to extract mobility patterns or attempting localization. 

Our motivation behind the use of CDR data can be resumed into its potential of reflecting human mobility characteristics and depicting the population movement and displacement in urban and rural areas. Therefore, our purpose is to try to transform CDR based trajectories to GPS like trajectories by estimating the exact location of the mobile users by relying only on the CDR data and coverage areas locations. The resulting outcome can be used for many localization applications for tracking purposes, information sharing or broadcasting, target marketing, fleet management, travel information systems, intelligent transportation systems, etc.
%----------------------------------------
\section{Related work}
Location information or positioning is important for many applications related to monitoring the activities in the telecommunication networks, navigation, mobility management, transportation. This importance comes from the fact that this type of knowledge can help in understanding human mobility patterns, displacement, and daily activities.

From this perspective, there is a lot of research done using GPS data as a source of positioning sensors. For example, in \cite{Dash2015} the authors collected GPS data and built an algorithm for extracting meaningful locations. Then, they predicted users' movement based on a mobility model created using the discovered displacement patterns. Of course, GPS data can be considered as a very good source of data for building mobility behavior models and localization; however, not all of the mobile phone users are happy to share this kind of information all the time, whenever and wherever they are. That's why CDR data can be a potential source of data for localization and mobility modeling \cite{Berry2000}.
Moreover, telecommunication data has been collected continuously for many years now, which makes this type of data available in massive amounts \cite{Kana2013}. Furthermore, by correlating the CDR data and the geographic locations of towers as mentioned before, the CDR data can be very useful for discovering and extracting mobility patterns \cite{Issac2012}. 

Therefore, we can distinguish three major approaches one is focusing on the spatial characteristics of movements and their influence on human mobility, using statistical techniques to depict the movement patterns or on the characteristics of the radio waves generated by the telecommunication antennas.
Concerning the spatial approaches, they are focused on the use of measures such as jump length, radius of gyration, or centroids \cite{Bar2008}. These features allow performing the extraction of meaningful locations since people have a tendency to stay in key locations and spend some time in those areas. In \cite{Issac2011} the authors used the same ideology and proposed an algorithm to identify significant locations using CDR data. This results demonstrated that the proposed approach was capable of finding home and work location with 88\% accuracy.  However, the method was not designed to discover other important locations that can occur during the movement between work and home. 

Most of the studies conducted regarding mobility are trajectory based where each user’s trajectories are traced, extracted, and studied from a behavioral point of view.  In the end, all these studies or research are trying to answer specific questions that are the key to innovation in localization and mobility patterns in both research and business applications.  The main or general question can be categorized into the following categories: displacement \cite{Huang2013}, movement patterns or models \cite{Krug2014}, \cite{Sun2014}, and human daily activities or resources \cite{Ton2015}, \cite{Khan2015}.

At this level, it is clear that many applications need not only the detection of behavioral patterns of mobility but also more accurate localization awareness using mobile networking. The statistical and probabilistic approaches have proven to be robust in providing a clear understanding the mobility patterns and insights discovery. For example, the authors in \cite{Hu2004} proposed a method based on the sequential Monte Carlo for exploiting mobility, improving the accuracy and precision of positioning. The technique was tested based on a simulation and it was demonstrated that it can provide accuracy in localization. However, the model used for representing the mobile network coverage does not take into account that the coverage zone can fluctuate and change shapes due to an irregularity of network transmissions.

In the same category, we a have the same method Sequential Monte Carlo (SMC) used in a different manner presented in \cite{Dil2006}, where the authors tried to enhance the application of SMC by making it robust regarding the localization estimation even if the range measurement error is high and unpredictable. The testing of the algorithm was done using a simulation and the results showed that the approach is capable of improving the localization range from 12\% to 49\% under a wide range of conditions. The model presented gave interesting results but there was no testing in real life situations or with different models using simulations.

Regarding the methods based on propagation time and signal strength \cite{H2015}. Signal strength approaches are focused on received signal strength indicators, which reflect the attenuation measurement of the signal in the assumed free space propagation of radio signals. However, the reality might be described as a free space propagation; therefore, it affects the triangulation method for localization. 

These phenomena influencing the signal are as follows: penetration, reflection, scattering, and diffraction. To overcome this problem, some propagation models have been proposed, such as software-defined radios model (SDR) \cite{Blo2013} or cost231-wolfish-Ikegami model \cite{Wal1988}. Whereas, propagation time-based methods rely simply on the time measurement, which makes their accuracy depending on the number of measurements acquired and on the geographical environment (urban areas are a big obstacle to its performance).
Furthermore, some other researchers investigated localization using mobile users, but for an indoor case. For example, the work in \cite{Lee2008} proposed an algorithm that enables the localization of mobile users by centering on trajectory matching in indoor settings. The mobility of the users was modeled by learning from the signal strength from their phones' historical data. The results showed that the localization had an average accuracy of 1.3 m. However, the environment was controlled as well as the measurement, which makes the results non-realistic with regard to the sparseness in time and space of real mobile data.

%----------------------------------------
\section{Data description and problem statement }
\subsection{Nature of the data}
\begin{figure}[!ht]
\centering
\includegraphics[width=0.6\textwidth]{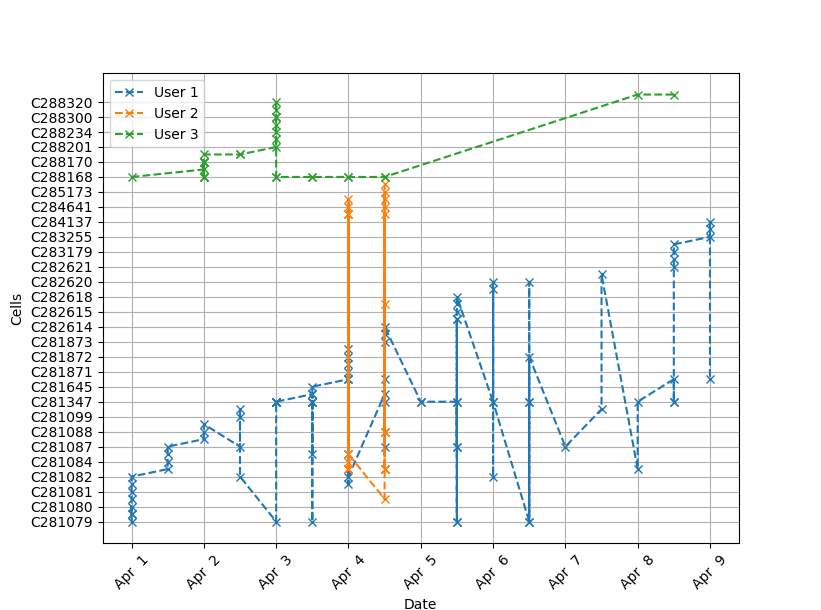}
\caption{Users' trajectory extracted from CDR data and their triggered time - each color represent a different user}
\label{fig:temp}
\end{figure}

\begin{figure}[!ht]
\centering
\includegraphics[width=0.6\textwidth]{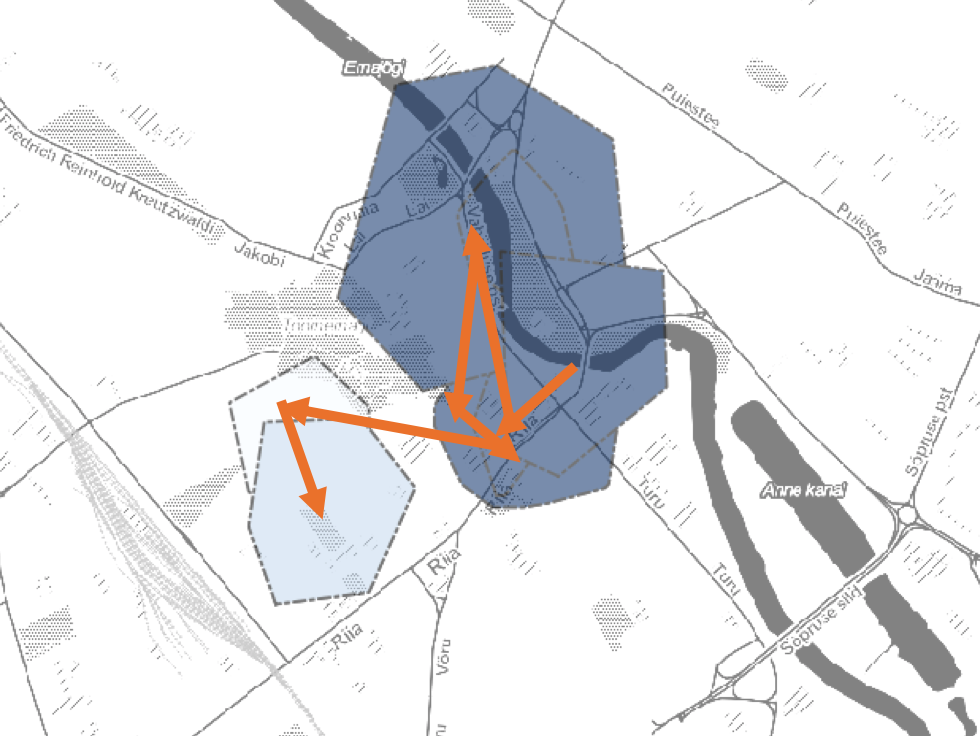}
\caption{User's Cell based trajectory extracted from combined CDR data and coverage areas locations - polygons are the coverage areas and the arrows are the chronological order of the event occurrence}
\label{fig:spat}
\end{figure}

Call detail records or CDR is a set of information about telecoms' transactions that the operator uses to generate billing. The record is generated at the moment of an explicit phone usage, such as calling, SMS or packet data. The collected records contain the following information (Table \ref{tab:cdr}): 
\begin{table}[!ht]
\renewcommand{\arraystretch}{1.3}
\caption{Attributes Contained in the CDR Data}
\label{tab:cdr}
\centering
\begin{tabular}{ll}
\hline
Attributes & Description\\
\hline
 IMSI & refers to the international mobile subscriber\\& identity\\
%\hline
 IMEI & refers to the international mobile equipment \\& identity \\
%\hline
 CellID & refers to the id of the transmitter providing \\& the coverage\\
%\hline
 timestamp &  time when the event occurred \\
%\hline
 Event & refers to the type of the event that triggered \\& the registration of the record  such as:\\& call, sms, browsing, etc. \\
\hline
\end{tabular}
\end{table}
The CDR data gives us temporal information about the event triggered by mobile devices and to which coverage area (CellID) the connection was established. In order to add the spatial aspect to the CDR data, we add the locations and the geographical shape of the coverage area using the registered CellID. Hence, CDR data is spatio-temporal data about the mobile users' daily activities. 
In general, the CDR data is very sparse in time and space. This can be easily observed by checking figure \ref{fig:temp}, which illustrates the trajectory of a user in time. We can see that the records are very sparse in time. Furthermore, the sparseness of the data affects also the locations, as it is visible in the figure \ref{fig:spat}.

From this first glimpse, we can clearly spot the sparseness in time and space of the registered events. The CDR data is very sparse in time and also from the geographical point of view the locations of the coverage areas are very distance from each other.

\subsection{Problem statement}

Based on the nature of the data described above, our objective is to find out a way to estimate the exact localization or the positioning of mobile users inside the mobile network coverage. Therefore, we have to optimize the coverage areas, make a classification of the movement type and at the same time estimate the exact position of the user. 

\begin{figure}[!ht]
\centering
\includegraphics[width=0.6\textwidth]{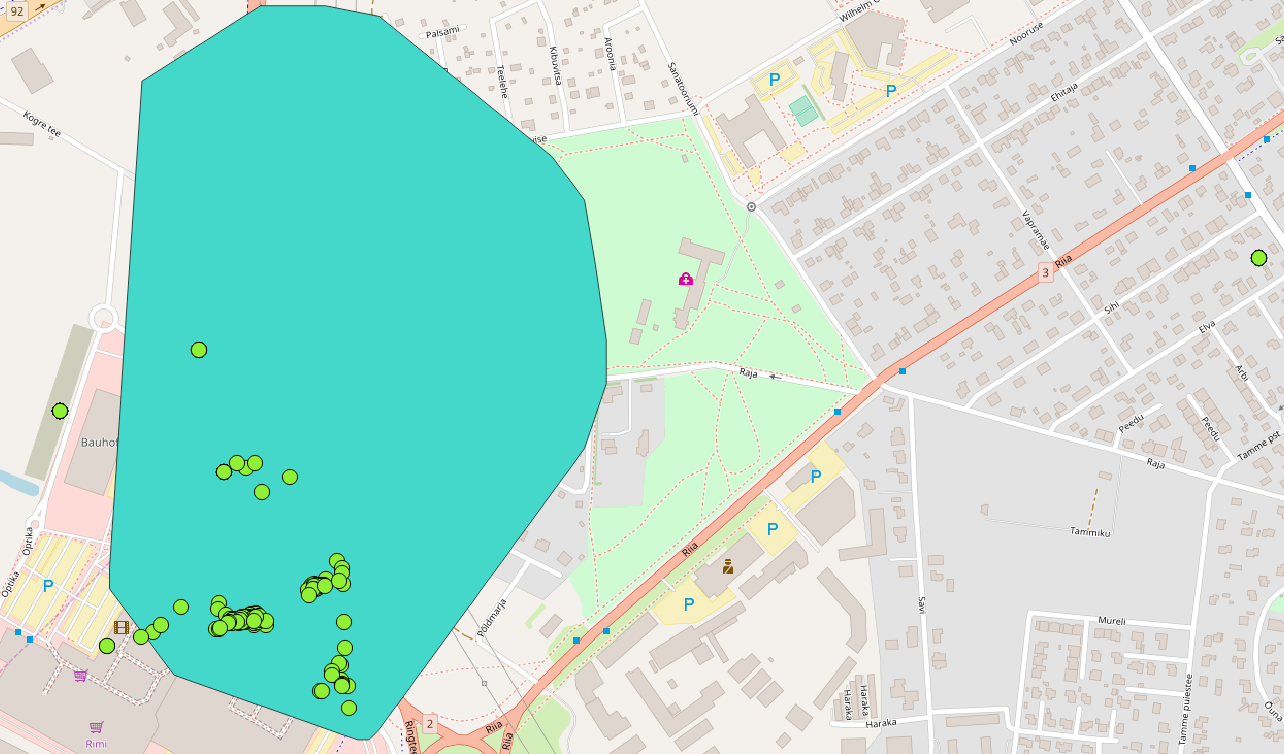}
\caption{Illustration of wrong representation of the coverage areas due to radio signal fluctuation phenomena - Coverage area generated using Voronoi (polygon) and GPS locations of the device connected to it (dots)}
\label{fig:cov}
\end{figure}

For this reason, the proposed algorithm is capable of detecting the type of moment (moving or staying) between successive cells' records within the trajectory and also of estimating the localization of the mobile user inside the coverage areas or zones. The movement episode detection, that the adaptive Kalman filter performs, is based on the movement models used in the algorithm process. Then, the Kalman filter algorithm makes an estimation of the exact position of the mobile phone user inside the coverage area. Finally, the estimated position is map-matched to the right road segment if the movement episode detected is \textit{moving} or to the correct building if the movement episode detected is \textit{staying}. Moreover, we noticed that sometimes the connection is established with a specific coverage area even if the mobile device is located outside the coverage zone (figure \ref{fig:cov}). This phenomena is due to the signal strength fluctuation of the radio waves \cite{Doh2014}. Therefore, we will also introduce a technique to optimize the coverage area in such a manner to reflect the real propagation of the coverage area. 

%----------------------------------------
\section{Adopted approach}
Estimating the precise location of mobile users within the coverage areas using only the CDR data is a very challenging task. Therefore, our methodology is inspired and based on previous work \cite{lind2017new}, where first the approach starts by optimizing the coverage areas. This task is performed using a data-driven model that will be explained in the following subsection. Then, we will apply our switching Kalman filter to the detected type of movement registered in the CDR data and also estimate the location of mobile devices within the mobile network coverage area . Finally, we will introduce map-matching techniques to map the estimated location to roads or building based on the outcome of movement models. 

\subsection{Coverage optimization}
As mentioned before, the coverage area provided by the mobile operator and estimated using Voronoi method needs to be corrected. Hence, we collected GPS data as illustrated in figure \ref{fig:cov} to verify how much of the coverage is wrong. From this perspective, we designed our approach for enhancing the representation of the coverage area by performing the following steps:

\begin{enumerate}
\item First, we create a circle around the polygon in such manner that the antenna location and the azimuth of the signal propagation are taken into account. 
\item Second, we execute a mathematical optimization by minimizing a penalty function $\mathsf{f(p)}$, where $p$ refers to radius of extension, based on the observations (in our case GPS data) that are applied to the radius of the circle.
\end{enumerate}

The defined function $\mathsf{f(p)}$ penalizes large distances $\mathsf{d(x,r,y)}$ between the coverage zone (cell area) $\mathsf{(x,r)}$, $\mathsf{x}$ refers to the center and $\mathsf{r}$ to the radius, and the GPS coordinates $\mathsf{y}$ when the events was triggered. In addition, we correct the large coverage areas that are not realistic to the GPS data by reducing the size of the coverage.

Therefore, in order to formulate our mathematical model let's consider $\mathsf{C}$ to be the set of all pairs of a cell index and its associated GPS coordinates $\mathsf{(j,y)}$. Our penalty function can be then expressed as follows: 

\begin{eqnarray}
f(\mathbf{p}) & = & \sum_{i}p_{i}^{2}+w\sum_{(j,\mathbf{y})\in \mathcal{C}}\left[\min\left(0,d(\mathbf{x}_{j},r_{j},\mathbf{y})\right)\right]^{2},
\end{eqnarray}

We defined $w=10$ as the weight for non-coverage penalty based on heuristic observations, $p_{i}$ as radius extensions, and $d(\mathbf{x},r,\mathbf{y})=r-\left|\mathbf{x}-\mathbf{y}\right|$. In addition, our implementation of the L-BFGS-B algorithm with the intention of \cite{Mora2011} minimizing the coverage function was based on   \emph{scipy} library.

\subsection{Mobility classification and localization}
The classical Kalman filter allows the use of only a single transition matrix at each step $t={1,..,T}$. Thus, it is a dynamic linear system defined as follows:
\begin{eqnarray}
\mathbf{x}_{t} & = & F\mathbf{x}_{t-1}+\mathbf{q}_{t}\\
\mathbf{y}_{t} & = & H\mathbf{x}_{t}+\mathbf{r}_{t}\:,
\end{eqnarray}

Where, $\mathbf{x}_{t}$ and $\mathbf{y}_{t}$ represent the hidden state and observed evidence respectively. Furthermore, $F$ is a transition matrix and $H$ is an observation matrix. The last variables in both equations are random Gaussian noises for transition $\mathbf{q}_{t}\sim\mathcal{N}(0,Q_{t})$ and for observation $\mathbf{r}_{t}\sim\mathcal{N}(0,R_{t})$.

In general, the Kalman filter represents the belief states $\mathbf{x}_{t}$ of using continuous random variables $\mathbf{X}_{t}$ with Gaussian probability distributions $\mathbf{x}_{t}\sim P(\mathbf{X}_{t}=\mathbf{x})=\mathcal{N}(\mathbf{x};\mu_{t},\sigma_{t})$. In addition, the inference algorithms can be used to calculate the probability distribution of $\mathbf{X}_{t}$ from the up-to-date evidence $\mathbf{y}_{1:t}$, which is \emph{filtering}, or all the evidence $\mathbf{y}_{1:T}$ - known as \emph{smoothing}.

The Kalman filter is known to be very good in estimating or predicting a state following one behavioral model. Whereas, human mobility in real life has many behavioral types and it can involve multiple mobility models at each step in a trajectory. Therefore, we will use adaptive Kalman filter \cite{Xiao2016}, where we will introduce discrete random variables to the algorithm. 
The adopted discrete variable $S_t$ will define the chosen model used in the transition step at $t$. Next, the Kalman filter computes the probability of each model $\mathsf{M}$ at time $t$, given all the evidence (Formula 6 \& 7) and the probability distribution of the hidden state variable associated with each model (Formula 8).

\begin{eqnarray}
M_{t|t}(i) & = & P(S_{t}=i|\mathbf{y}_{1:t})\\
M_{t|T}(i) & = & P(S_{t}=i|\mathbf{y}_{1:T})
\end{eqnarray}
\begin{eqnarray}
P(\mathbf{X}_{t}|S_{t}=i,\mathbf{y}_{1:\tau})=\mathcal{N}(\mu_{t|\tau}^{i},\Sigma_{t|\tau}^{i})\quad\tau\in\{t,T\}
\end{eqnarray}

In addition, the consolidated belief state of hidden variable $\mathbf{X}_{t}$ at time $t$ is expressed by combining all the Gaussians models and the probabilities of the models as follows: 

\begin{equation}
P(\mathbf{X}_{t}|\mathbf{y}_{1:\tau})=\sum_{i}M_{t|\tau}(i)\cdot P(\mathbf{X}_{t}|S_{t}=i,\mathbf{y}_{1:\tau})\end{equation}

Finally, it is necessary to define the model transition probability matrix $T_z(i,j)=P(S_{t}=j|S_{t-1}=i)$. We define the transition probability
from $S_{t-1}=i$ to $S_{t}=j$ with the higher chance to stay in the same model: 

\begin{equation}
T_z(i,j)=\begin{cases}
0.8 & \mbox{if\,}i=j\\
\frac{0.2}{Nu_{m}-1} & \mbox{otherwise}
\end{cases}
\end{equation}

where, $Nu_{m}$ is the number of models. Furthermore, one of the issues with KF is the exponential growth of the belief state due to the multiplication between the number of Gaussians and the number of models at each step. For this reason, filtering and smoothing are a necessity in the process and they are computed using predefined models of behavior.

%%%%%%%
After defining and explaining the main component in our adaptive Kalman filter, we will explain how the mobility models are defined and integrated in the algorithm.  We begun by considering location $\bar{\mathbf{x}}_{t}$ and velocity $\mathcal{\vartheta}_{t}$
at time $t$ as hidden variables $\mathbf{x}_{t}=\left(\begin{array}{c}
\bar{\mathbf{x}}_{t}\\
\mathcal{\vartheta}_{t}
\end{array}\right)$ and ensuring that a \emph{moving} user`s coordinates and velocities
comply with the following equations: 

\begin{eqnarray}
\bar{\mathbf{x}}_{t} & = & \bar{\mathbf{x}}_{t-1}+\mathcal{\vartheta}_{t-1}\delta t+\bar{\mathbf{q}}_{t}\\
\mathcal{\vartheta}_{t} & = & \mathcal{\vartheta}_{t-1}+\dot{\mathbf{q}}_{t}
\end{eqnarray}

where $\delta t$ is the time difference from the previous event resulting from the Bayes Network approach and $\left(\begin{array}{c}
\bar{\mathbf{q}}_{t}\\
\dot{\mathbf{q}}_{t}
\end{array}\right)$ are noise.
In conclusion, the KF equations are as follows:

\begin{eqnarray}
\mathbf{x}_{t} & = & F\mathbf{x}_{t-1}+\mathcal{Q}_{t}\\
\mathbf{y}_{t} & = & H\mathbf{x}_{t}+\mathcal{R}_{t}\:,
\end{eqnarray}

Hence, for each model we have to define a transition matrix $F$
and a noise variance matrix $\mathcal{Q}_{t}\sim\mathcal{N}(0,\mathcal{Q}_t^{(M)})$. For instance, in case of a \emph{moving} user on plain 2D map, its transition matrix (Move model) is:

\begin{equation}
F^{(M)}=\left(\begin{array}{cccc}
1 & 0 & \delta t& 0\\
0 & 1 & 0 & \delta t\\
0 & 0 & 1 & 0\\
0 & 0 & 0 & 1
\end{array}\right)\label{eq:FMoving}
\end{equation}

On the other hand, a \emph{staying} user at the same location can be characterized by an identity matrix $F^{(S)}=I$ (Stay Model). 

\begin{equation}
F^{(S)}=\left(\begin{array}{cccc}
1 & 0 & 0& 0\\
0 & 1 & 0 & 0\\
0 & 0 & 1 & 0\\
0 & 0 & 0 & 1
\end{array}\right)
\end{equation}

At the same time, the observation model ($H$ and $R_t$) should reflect the location of the antenna that the user is connected to, and the coverage zone of the antenna expresses the observation error. During testing, all the "antennas" will have the same model:

\begin{equation}
H=\left(\begin{array}{cccc}
1 & 0 & 0 & 0\\
0 & 1 & 0 & 0
\end{array}\right)\quad R_t=\left(\begin{array}{cc}
1.2^{2} & 0\\
0 & 1.2^{2}
\end{array}\right)\:.
\end{equation}

\begin{figure}[!ht]
\centering
\includegraphics[width=0.4\textwidth]{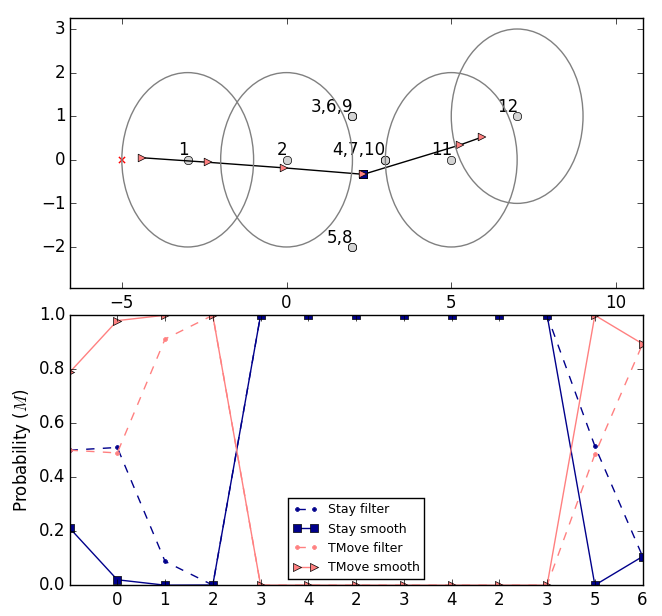}
\caption{Illustration of the algorithm steps in evaluating the Observations and computing the probabilities for Stay and Move models - The circle and dots represent the observed coverage areas and below the associated probabilities following the chronological order}
\label{res}
\end{figure}

As a consequence, the algorithm computes the probabilities $P(S_{t}=k|\mathbf{y}_{1:t})$ of each model $k$ at time $t$ and the probability distribution of the coordinate and the velocity $P(\mathcal{X}_{t}|S_{t}=k,\mathbf{y}_{1:t})$ for given up-to-date evidence $\mathbf{y}_{1:t}$. The figure \ref{res} illustrate exactly this process of computing the probability of each model based on the observed evidence in our case the triggered event within a coverage area. Then, the same process is applied to all the evidence $\mathbf{y}_{1:T}$, which gives smoothed results, that are more accurate and therefore are used in actual testing and validation.

%%%%%%
\subsection{Map-matching}
\begin{figure}[!ht]
\centering
\includegraphics[width=0.4\textwidth]{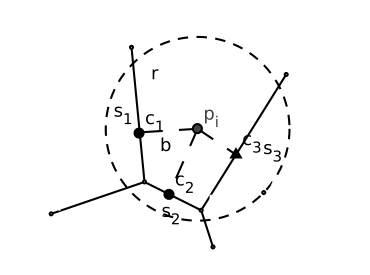}
\caption{Map Matching schemtic: $p_i$ denotes the point that has to be matched; $s_1$, $s_2$ and $s_3$ are the road segments found in radius $r$; $c_1$,$c_2$ and $c_3$ are the projections of the $p_i$. Projection $c_3$ is selected for matching $p_i$ as it is the closest to $p_i$.}
\label{fig:map_match}
\end{figure}

In order to improve the accuracy of the estimated positions, the corresponding points were matched to the closest road (Figure \ref{fig:map_match}). The applied Map-matching on the point $p_i$ is performed in two steps: first, the road segments set $S_{p_i,r}$ in certain radius $r$ around the $p_i$ are looked up to by applying the following formula :

\begin{equation}
S_{p_i,r} = \{\ (p_1,p_2)\ |\ d(p_1,p_i) <= r \lor d(p_2,p_i) <= r\ \}\ ;
\end{equation}
\label{form:roads_in_r}

Next for each found road segment $s_j$ a matched candidates points set $C_{p_i,r}$ are computed using an \textit{orthogonal projection} ($proj(,)$) of $p_i$ to $s_j$:

\begin{equation}
C_{p_i,r} = \{\ proj(p_i,s_j) | \forall s_j \in S_{p_i,r} \}\ ;
\end{equation}
\label{form:match_candidates}

Afterwards a distance $d(p_i ,c_j)$ between point $p_i$ and all its matched candidates $c_j$ is computed using \textit{haversine distance} \cite{Carl1957}, since we are dealing with a projection on a sphere (Earth). Finally minimal distant candidate is selected;

\begin{equation}
mm(p_i,r) = \{\ c | d(p_i,c) = min [ d(p_i,c_j), c_j \in C_{p_i,r} ] \ \}\ ;
\end{equation}
\label{form:point_match}

%----------------------------------------
\section{Experimental evaluation}

The evaluation of our method is executed by comparing it to GPS data collected from the field as a ground truth. The GPS data was collected using an application developed by us in order to add an annotation to the collected records. This way, we can evaluate the movement episode detection (Move or Stay) and also the estimated locations after and before the map-matching process. 
The data used for testing contains 649 CDR records provided by the mobile operator for six different users for almost a period of one month. Furthermore, during the campaign of collecting the data the users annotated their movement and we had 450 stay locations and 199 move locations. 

\subsection{Mobility episode detection}

\begin{figure}[!ht]
\centering
\includegraphics[width=0.5\textwidth]{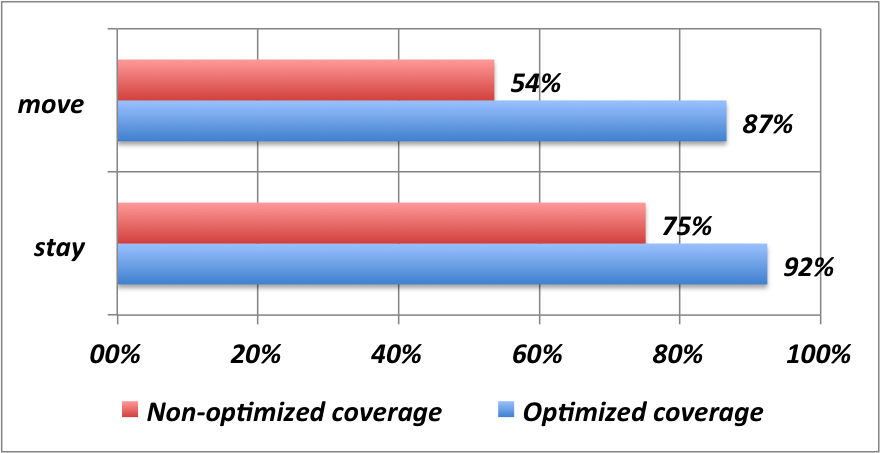}
\caption{Algorithm's performance in detecting movement episodes}
\label{fig:graph}
\end{figure}

By using the annotated GPS data we have an idea about the users whether they are staying in a specific location or moving. This information was used to assess the performance of our Kalman algorithm. The general performance can be visualized in figure \ref{fig:graph}, where we also tested out how the optimization of the coverage area influences the detection of movement episodes. The algorithm was capable of estimating stay locations with an accuracy of 75\% without introducing our method for optimizing the coverage area and an accuracy of 92\% with the proposed coverage optimization formula. The same increase happened to the detection of move locations from 54\% to 87\% of accuracy. It is clear that the coverage optimization approach has a positive impact on enhancing our detection and classification of the locations: a "move" location or a "stay" location. 

\subsection{Localization without map-matching}
\begin{figure}[!ht]
\centering
\includegraphics[width=0.5\textwidth]{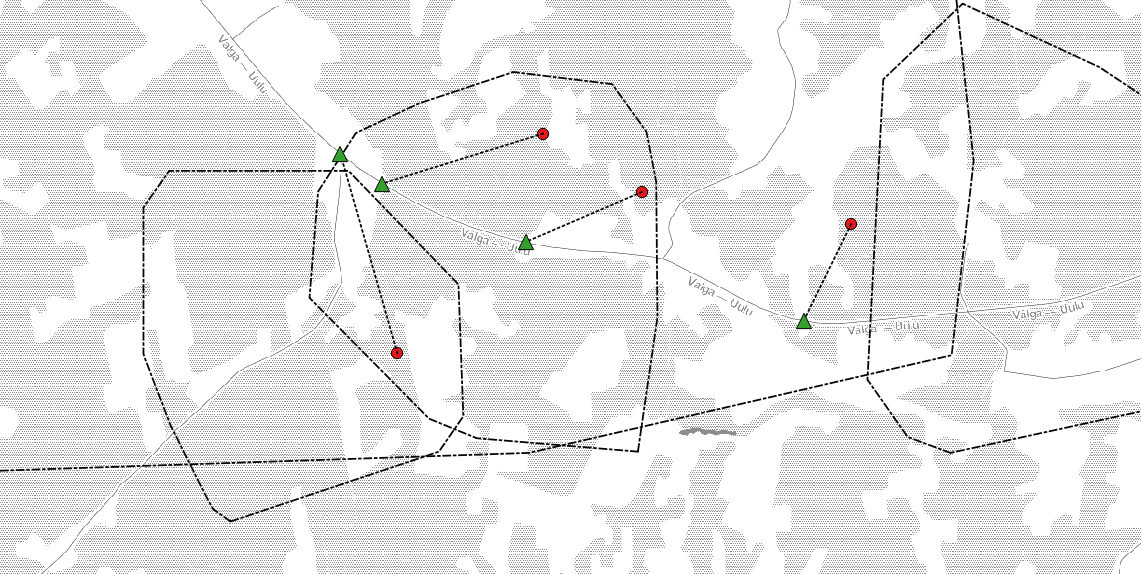}
\caption{The proposed model results without applying map-matching on the map: `` case of a traveling mobile user between two cities`` - small circles are the estimated positions; triangles are GPS locations; polygons are the coverage areas of the mobile networks }
\label{fig:trj}
\end{figure}

The next test is the main one since our interest lies in positioning mobile users within coverage areas using only CDR data with a precision similar to the GPS. Hence, we compared the estimated positions given by the adaptive Kalman filter to the real locations of the mobile users (GPS locations). As an illustration of our results, you can see in figure \ref{fig:trj} the case of the user traveling between two cities. The triangles represent his/her GPS data during the trip and the circles are the estimated positions using our method. In addition, every circle dot is linked to its appropriate ground truth GPS data by a segment. 

\begin{figure}[!ht]
\centering
\includegraphics[width=0.6\textwidth]{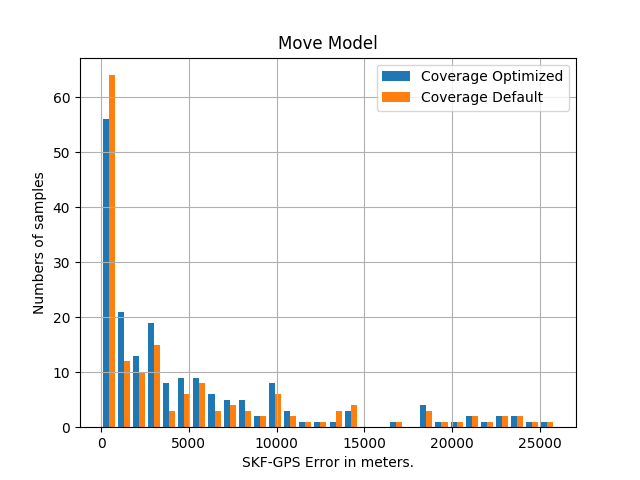}
\caption{Error distribution of the proposed algorithm for "Move" locations}
\label{fig:r1}
\end{figure}

In addition, for more clarity, we analyzed the estimation error through the comparison of the estimated locations and the real locations provided by the GPS data. The general view about the result with respect to "Move Locations", the location estimation error is illustrated in figure \ref{fig:r1}, where the error distribution is shown for both cases: estimation with or without applying any coverage optimization. The algorithm uses CDR data combined with coverage area (polygon and centroid) to estimate the location of the mobile device within the coverage areas and it is noticeable that the error distribution gets better by applying the optimization since it is more concentrated around zero error. Furthermore, the error distribution when using no coverage optimization has a mean of 4912.9 meters and a standard deviation of 6510.4 meters and when using optimization the mean is 4937 meters and the standard deviation is equal to 6083.2 meters It is clear that coverage optimization has an impact on the accuracy of the estimation of our proposed model.

\begin{figure}[!ht]
\centering
\includegraphics[width=0.6\textwidth]{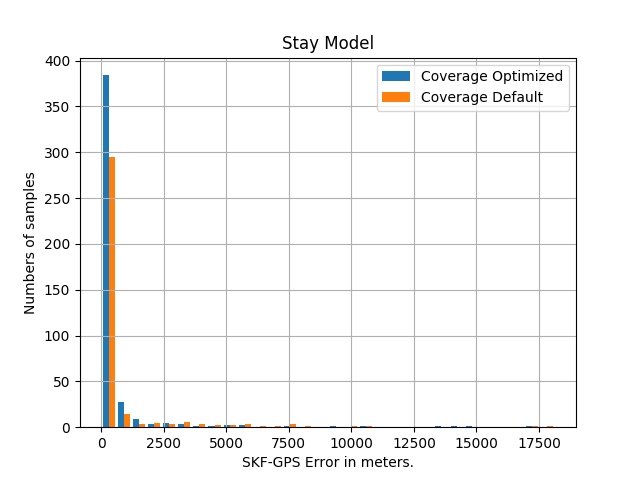}
\caption{Error distribution of the proposed algorithm for "Stay" locations}
\label{fig:r2}
\end{figure}

The same impact has been recorded for the "Stay Locations" since the error distribution  with a mean of 637 meters and a standard deviation of 2008.1 meters in the case of no coverage optimization and it got more distributed around zero by applying coverage optimization which is reflected in a mean of 476.4 meters and standard deviation of 1739.2 meters (Figure \ref{fig:r2}).

\subsection{Localization with map-matching}
\begin{figure}[!ht]
\centering
\includegraphics[width=0.5\textwidth]{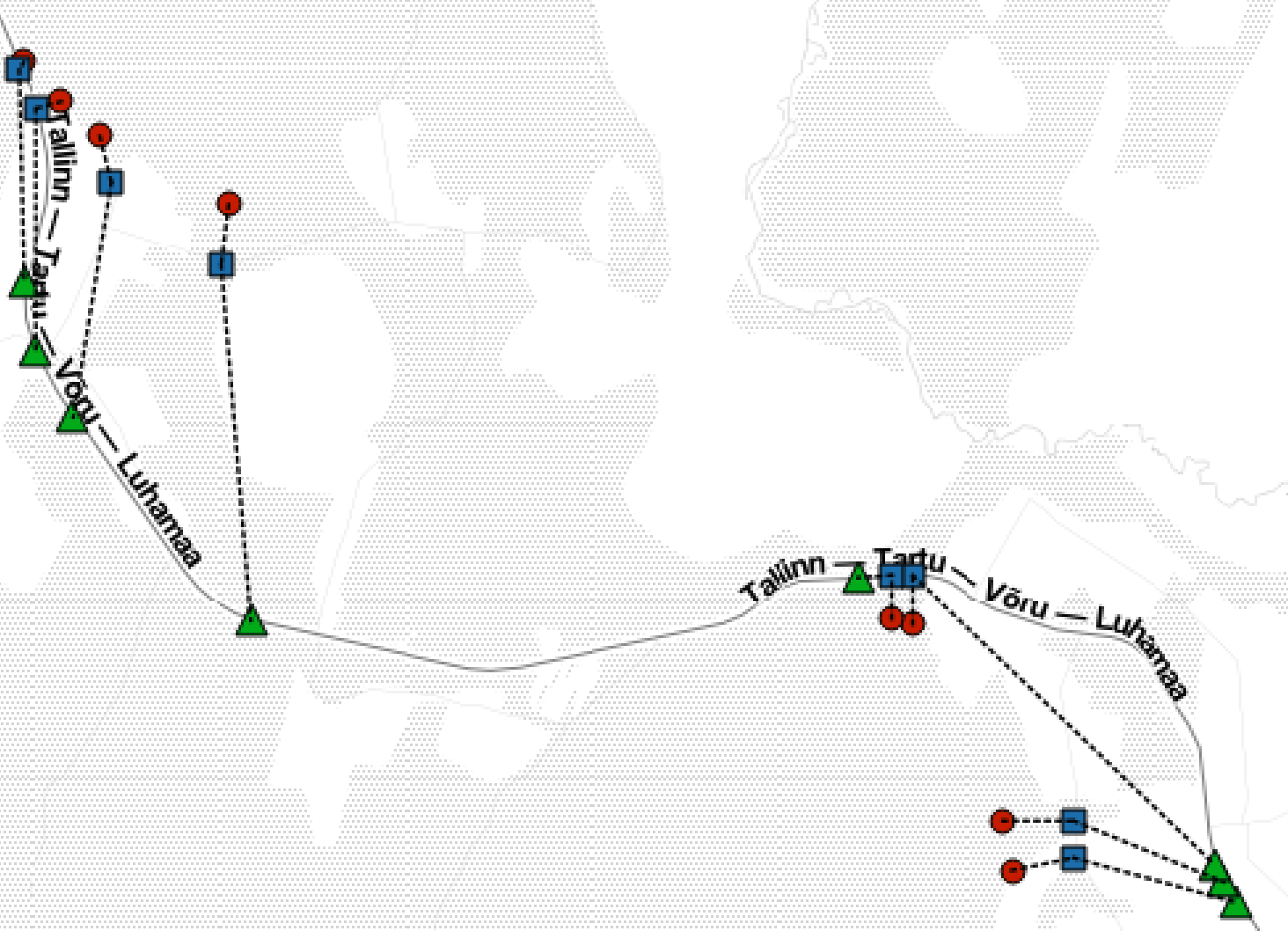}
\caption{Validating Map-Matched results of Kalman filter (no coverage optimization); red circles - estimated position, blue squares - map-matched position, green triangles - actual position}
\label{fig:skf-mm-gps-noopt}
\end{figure}

\begin{figure}[!ht]
\centering
\includegraphics[width=0.5\textwidth]{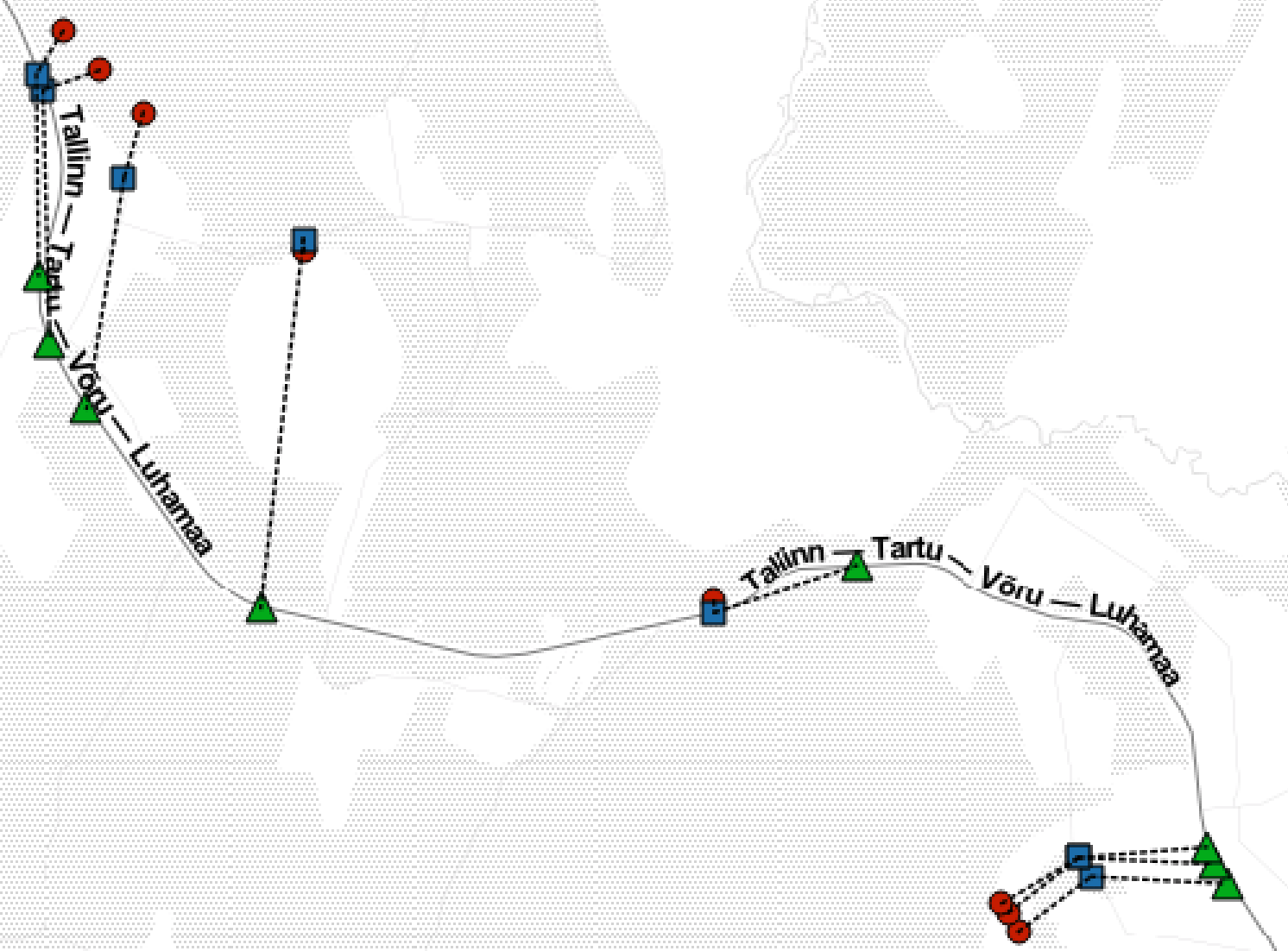}
\caption{Validating Map-Matched results of Kalman filter (using coverage optimization); red circles - estimated position, blue squares - map-matched position, green triangles - actual position}
\label{fig:skf-mm-gps-opt}
\end{figure}

Finally, the last test is about the impact of map-matching on estimating the position of mobile users. As explained in section four, the map matching takes as an input the estimated position and it is mapping it to the closest road if the model was a move model and to the closest building if the model was a stay model. The outcome of this process of evaluation using root mean square error (RMSE) is illustrated in table \ref{table3}, where we applied map-matching on the estimated position without coverage optimization (Figure \ref{fig:skf-mm-gps-noopt}) and with coverage optimization  (Figure \ref{fig:skf-mm-gps-opt}) . The results show that the map-matching has a good impact on the "Move location" since the error decreased and also it is clear that the coverage optimization has a positive impact on reducing the error. however, introducing the map-matching for "stay location" has impacted it negatively. Therefore, from this analysis, we can conclude that the best model for estimating "Stay locations" is by introducing only the coverage optimization and for "Move locations" introducing the coverage optimization and the map-matching has a good impact on reducing the error. 

 \begin{table}[!ht]
 % increase table row spacing, adjust to taste
 \renewcommand{\arraystretch}{.2}
 \caption{The RMSE evaluation in meters of the proposed algorithm for positioning with Coverage Optimization (Opt) and Map-Matching (MM)}
 \label{table3}
 \centering
 % Some packages, such as MDW tools, offer better commands for making tables
 % than the plain LaTeX2e tabular which is used here.
 \begin{tabular}{ccccc}
 \hline
 Model & No-opt & Opt  & No-Opt+MM & Opt+MM \\
 \hline
 \hline
 Stay  &2106.8 & 1803.3 & 2788.1 & 2402.6\\
 locations & &  & & \\
 \hline
 Move & 8156.1 & 7834.5  & 4712.1 & 3344.4\\
 locations &  &   &  & \\
 %\hline

 \hline
 \end{tabular}
 \end{table}

%==================================================
\section{Conclusion}

The emerging technologies related to mobile data especially CDR data has great potential for mobility and transportation applications. However, it presents some challenges due to its spatio-temporal characteristics and sparseness. Therefore, in this article, we introduced a new model to refine the positioning accuracy of mobile devices using only CDR data and coverage areas locations. The adopted method has three steps: first, we discovered which model of movement (\textit{Move, Stay}) is associated with the coverage areas where the mobile device was connected using a switching Kalman filter. Then, simultaneously we estimated the location or the position of the device. Finally, we applied map-matching to bring the positioning to the right road segment. The results are very encouraging and the approach performs with less error in the cases of "Move model" by adding coverage optimization and map-matching and for "Stay model" the optimization was enough to reduce the errors; nevertheless, there is some enhancement that can be done at the level of data preprocessing, movement models and map matching. For example by introducing more sophisticated movement model based on reinforced learning and a more sophisticated map matching for matching the estimated locations.

\bibliographystyle{unsrt}  
%\bibliography{references}  %%% Remove comment to use the external .bib file (using bibtex).
%%% and comment out the ``thebibliography'' section.

%%% Comment out this section when you \bibliography{references} is enabled.

\end{document}